\documentclass[12pt]{article}
\usepackage{times}
\usepackage{graphicx}
\usepackage{array, ltablex, multirow}
\usepackage[numbers,square,sort&compress]{natbib}
\usepackage{float}
\usepackage{authblk}
\usepackage{xcolor}

\date{}

\begin{document}
\title{Revealing the 3D Structure of Graphene Defects}

\author{Christoph Hofer}
\author{Christian Kramberger}
\author{Mohammad Reza Ahmadpour Monazam}
\author{Clemens Mangler}
\author{Andreas Mittelberger}
\author{Giacomo Argentero}
\author{Jani Kotakoski}
\author{Jannik C. Meyer}

\affil[1]{Faculty of Physics, University of Vienna, Boltzmanngasse 5, A-1090, Vienna, Austria}

\maketitle

\begin{abstract}

We demonstrate insights into the three-dimensional structure of defects in graphene, in particular grain boundaries, obtained via a new approach from two transmission electron microscopy images recorded at different angles. The structure is obtained through an optimization process where both the atomic positions as well as the simulated imaging parameters are iteratively changed until the best possible match to the experimental images is found. We first demonstrate that this method works using an embedded defect in graphene that allows direct comparison to the computationally predicted three-dimensional shape.
We then applied the method to a set of grain boundary structures with misorientation angles nearly spanning the whole available range ($2.6-29.8^\circ$). The measured height variations at the boundaries reveal a strong correlation with the misorientation angle with lower angles resulting in stronger corrugation and larger kink angles. Our results allow for the first time a direct comparison with theoretical predictions for the corrugation at grain boundaries and we show that the measured kink angles are significantly smaller than the largest predicted ones.

\end{abstract}


Identifying the position of every atom in a sample is the ultimate goal of structural characterization. Although transmission electron microscopy (TEM) has already reached the spatial resolution to allow resolving all atomic distances\cite{Colliex2014,Krivanek1999}, it only provides two-dimensional (2D) projections of the sample regardless of their actual three-dimensional (3D) shape. While computer tomography can retrieve the 3D structure from a set of 2D projections, down to atomic resolution \cite{Bals2014,Miao2016,Weyland2004,Xu2015,Yang2017}, it requires high electron doses which is problematic for structures susceptible to electron-beam-induced structural changes. This is because in absence of any additional information, the number of projections required to obtain a uniform resolution in all dimensions is approximately the sample size divided by the resolution~\cite{Kak1988}, which reaches typically tens or hundreds of projections for bulk samples.

Defects in graphene, the 2D allotrope of carbon, are expected to corrugate the structure. Such corrugations have been studied previously through simulations~\cite{Yazyev2010,Liu2011,Kotakoski2014,Shekhawat2016} and their existence has also been indirectly inferred from high resolution TEM images \cite{Warner2013}\cite{Lehtinen2015}.
However, since graphene defects frequently change their configuration under electron irradiation even at moderate acceleration voltages~\cite{Kotakoski2011,Meyer2012,Susi2016,Kotakoski2014a,Susi2014,Kurasch2012}, recording an entire tomographic series to image the 3D structure of graphene defects at atomic resolution would be very challenging. The 3D structure of defect free graphene has been analyzed on the basis a of defocus series ~\cite{Dyck2012} and the structure of clustered divacancies has been extracted from atom contrast variations in a singe image \cite{Chen2017}. However, this approach requires that the intensity of each atom can be measured without being affected in any way by the intensity of the neighboring atoms, which is difficult to avoid in presence of residual aberrations, finite resolution, and very short projected distances in non-flat structures. The polynomial fit of the atom positions as done in Ref \cite{Chen2017} relaxes this requirement, but then it also does not reveal the position of individual atoms, but only averages of local height. In this way, it is not suited for the analysis of structures with sharp kinks or significant height differences between neighboring atoms, as revealed in this work. Our approach does not introduce such geometrical constraints and the results show that indeed the 3D configurations are more complex than the smooth height variations around defects that could be revealed previously. 

Here, we show that it is possible to obtain the 3D shape of defected graphene directly already from two experimental images obtained at different tilt angles. We first demonstrate our approach with an embedded graphene defect, for which the corrugated structure obtained from the experimental data can be directly compared with the one obtained computationally through energy optimization. Then, we move on to study the corrugations caused by grain boundaries in polycrystalline graphene. Importantly, both grain boundaries themselves as well as corrugations even in the absense of defects have been shown to significantly influence the properties of graphene~\cite{Rossi2015,Atanasov2010}, making this an important subject to study.



\begin{figure}

\includegraphics[width=\textwidth]{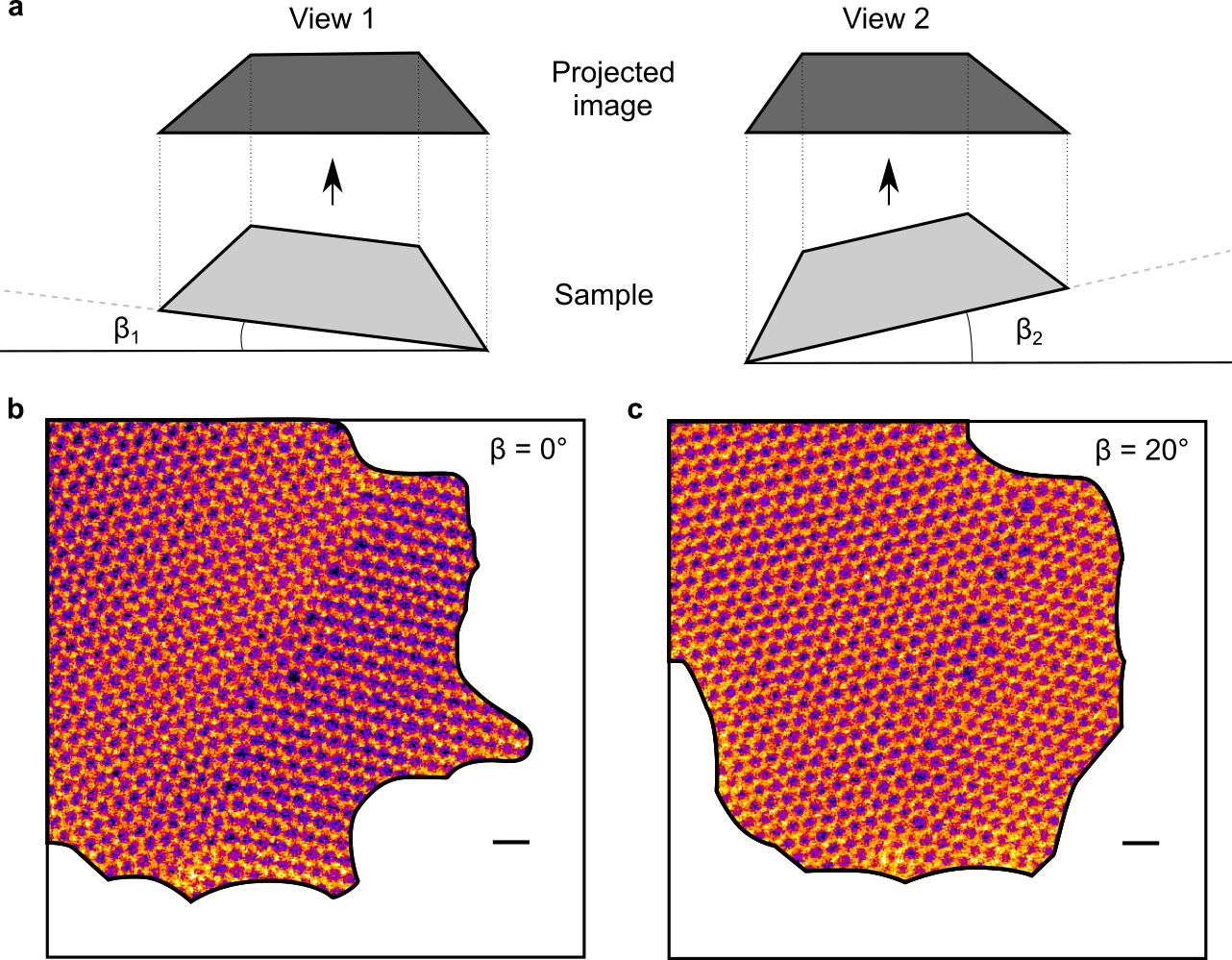}

    \caption{(a) Schematic illustration showing the sample at two different tilts ($\beta_1$ and $\beta_2$) that result in two different views of the sample (View 1 and View 2). (b) Filtered STEM-MAADF image of a graphene grain boundary at nominally zero sample tilt ($\beta = 0^\circ$). (c) STEM-MAADF image of the same grain boundary at a nominal tilt of ca. $\beta = 20^\circ$. The white areas correspond to contaminated areas that have been cut out from the images. Scale bars are 0.5 nm. The experimental images have been processed with a high pass filter to remove long range intensity variations. Raw images are shown in the supplementary information.}
\label{first}

\end{figure}

We start our experiment by looking for a defect in graphene grown via chemical vapor deposition (CVD; see Methods for details) using scanning transmission electron microscopy (STEM) medium-angle annular dark field (MAADF) imaging. In order to avoid electron irradiation-induced changes in the atomic structure~\cite{Kotakoski2011a}, the electron dose at the defect is minimized by recording as few atomic-resolution images as possible of the area of interest. After a defect is found and one atomic resolution image is acquired, we record a few images of the surrounding area at larger fields of view in order to find the same location after the sample has been tilted. While tilting, we track the sample to stay in the vicinity of the defect, and when the necessary tilt angle has been reached, we zoom in again and record the second atomic-resolution exposure. Even with this approach, the atomic structure at the defect often changes between the two recorded atomic-resolution images. However,  it is also possible to obtain pairs of images of different defect structures where the atomic structure remained unchanged. Areas covered by contamination in each image have been masked in order not to confuse the reconstruction algorithm and the images have been high-pass filtered. An example is shown in Figure \ref{first}.

\begin{figure}
\center \includegraphics[width=1\textwidth]{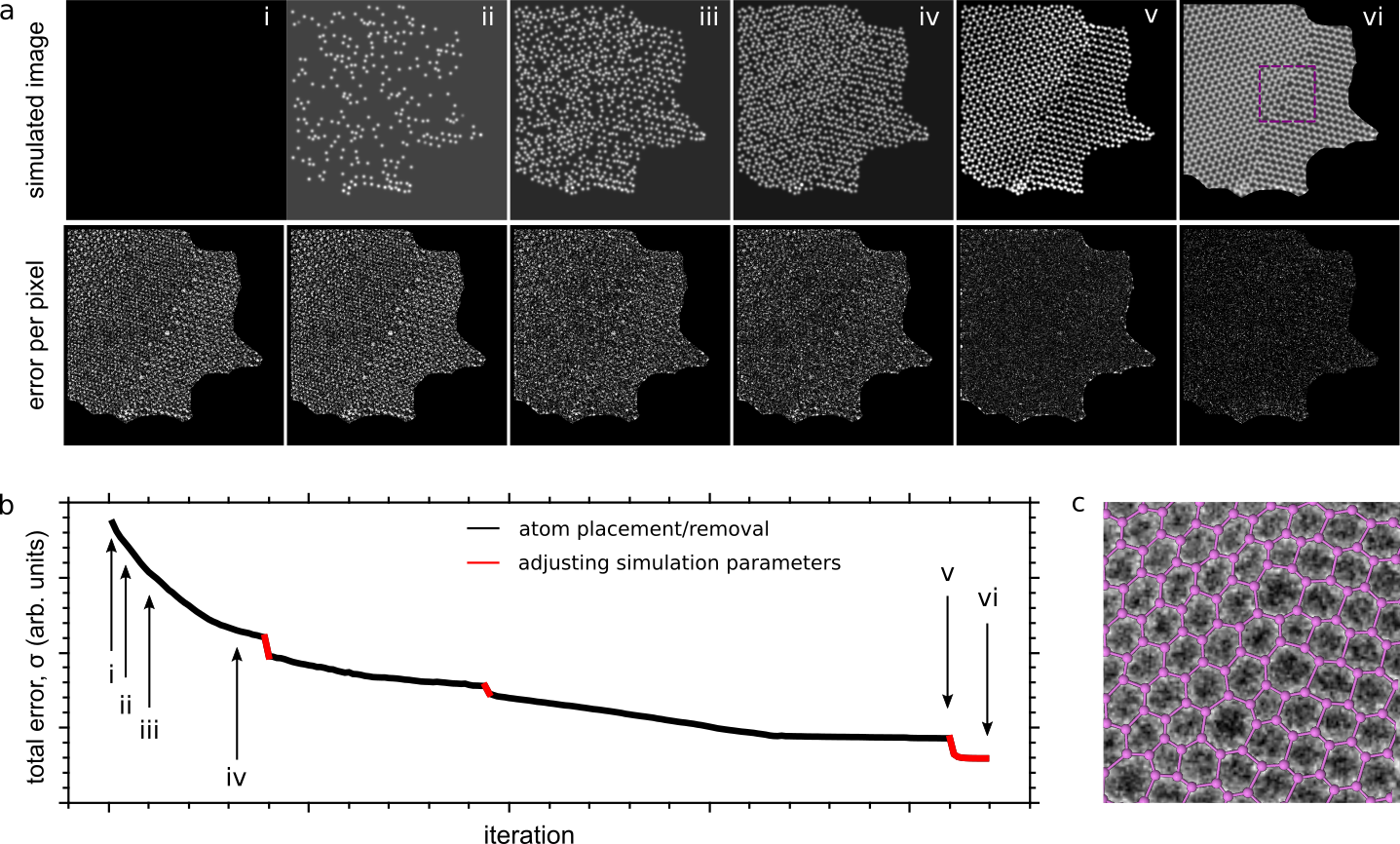}{}{}
    \caption{Error minimization through the optimization algorithm for one image. (a) Simulated image and the calculated error for each image pixel at six different stages (i, ii, iii, iv, v, vi) during the process. The corresponding optimization steps are marked with arrows in panel (b). (b) The calculated error $\sigma$ as a function of the number of iterations. The black data points correspond to the optimization of the atomic structure and the red ones to the optimization of the aberration coefficients used in the image simulation. (c) Overlay of the resulting 2D model and the experimental image, shown for a small section as indicated in the last frame of panel a.}
\label{twoD} 
\end{figure}

The first task for our reconstruction algorithm is the identification of individual atoms within each of the experimental images. This is achieved through an iterative process, where a model structure is compared at each step with the experimental image through image simulation (see Methods). Initially, the model contains no atoms. At each step, an atom is either added, its position is adjusted, or it is removed from the model. This is carried out by selecting a random position and either adding an atom there, or if there is an atom within a distance of $r_{cut} \leq 0.5$~{\AA} moving this atom there. If after this two atoms are too close to each other (within $r_{nbr} \leq 1$~{\AA}), the atom pair is replaced by just one atom. After each adjustment, a simulated STEM-MAADF image is created based on the model, and the difference between the simulated image and the experimental image is calculated as $\sigma = \sum_{i=1}^{N}(I^{exp}_i-I^{sim}_i)^2$, where the sum runs over all $N$ pixels in the image and $I^{exp}_i$ and $I^{sim}_i$ are the intensities of pixel $i$ in the experimental and the simulated image, respectively. At each step, the change in the structure is only accepted if it results in reducing $\sigma$ from the previous step. Since the match between the experimental and the simulated images depends not only on the exact atomic structure but also on the exact electron aberrations during the experiment (which can change between two images and need to be used as input parameters for the image simulation), they are also adjusted with a similar stochastic process. An image sequence showing one optimization process is shown in the Supplemental Material (Video 1) and the evolution of $\sigma$ as a function of the number of steps is shown in Figure~\ref{twoD}. At the end of this optimization process, $\sigma$ approaches a value close to zero and the difference image is dominated by noise.

Next, the topology needs to be established in order to allow identifying the same atom in each of the model structures. In our approach, this process is automated through the implementation of just a few rules. Firstly, two neighboring atoms need to be close enough (within 2~{\AA}) to allow bonding. Secondly, no more than three neighbous are allowed for each individual atom. Finally, the formation of three-membered carbon rings is prohibited. These rules are based on the obsersavation of previous atomic-resolution studies of the structure of $sp^2$-bonded defective carbon networks~\cite{Kotakoski2011, Eder2014} and appear to work extremely well. The agreement is easily confirmed visually by comparing the experimental image to the established network (see Figure~\ref{twoD}c) . Subsequently, identifying the same atoms in each of the images is possible based on their location in the network.

\begin{figure}
\center \includegraphics[width=\textwidth]{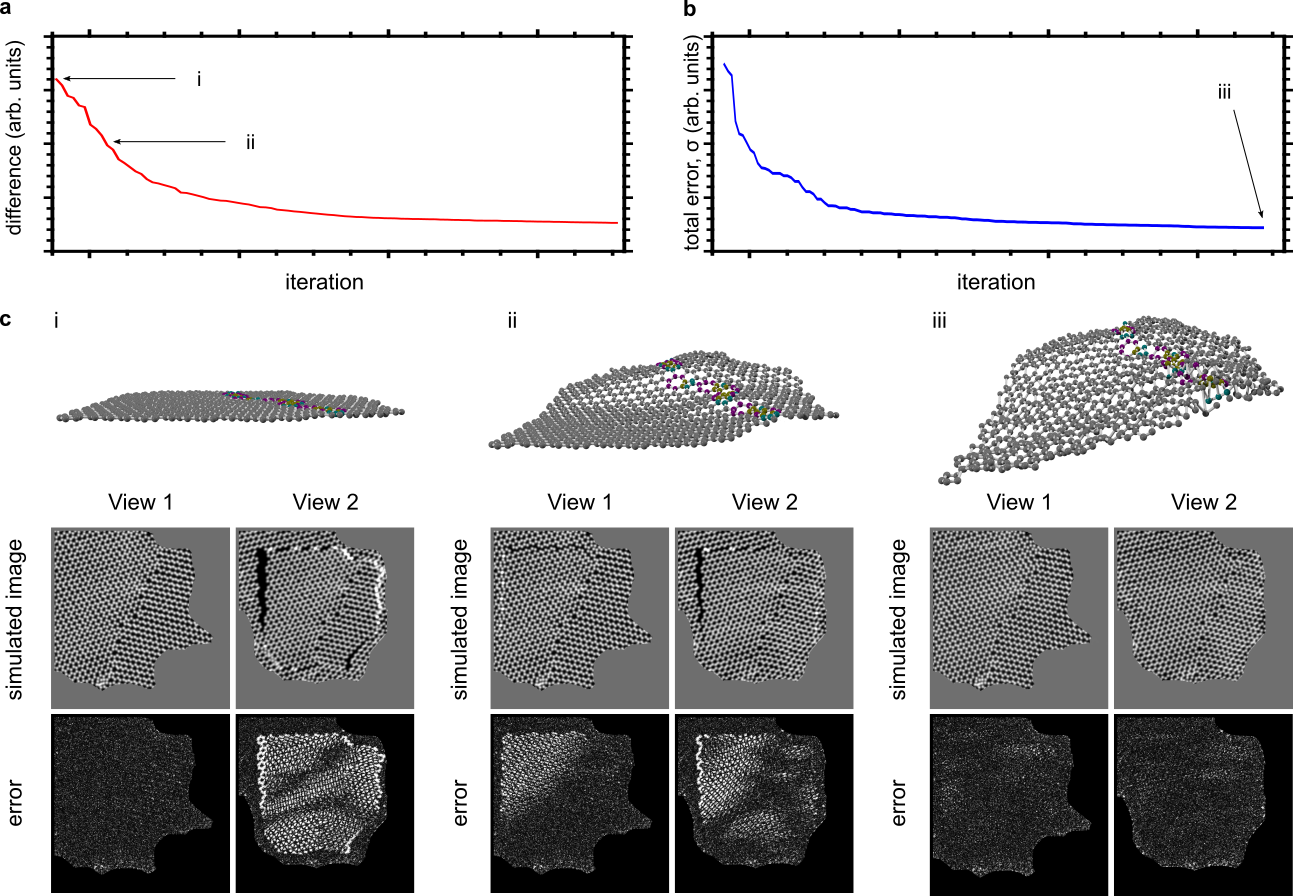}{}{}
    \caption{Evolution during the 3D optimization. (a) Difference between the projected atomic positions of the optimized 3D model and those in the flat models obtained during the previous optimization step. (b) Total error in the simulated STEM-MAADF images based on the 3D model, as compared to the experimental images. (c) Perspective views of the atomic structures (atoms of non-hexagonal rings in the grain boundary highlighted by color), simulated STEM-MAADF images for both tilt angles (View 1 and View 2), and the error for each pixel as compared to the experimental images. The area included in the 3D optimization is marked by the yellow dashed lines overlaid on the simulated images. The corresponding iteration steps for each case (i, ii, and iii) are marked in panels (a) and (b). 
		  }
\label{3D} 
\end{figure}

After the atoms have been identified, those that appear in both images (excluding image edges) are used as the basis for a new model which will be further optimized, now including also the third dimension. One of the 2D models is chosen arbitrarily for the initial positions of the atoms, and the optimization is continued considering both experimental views simultaneously (taking into account the tilt between the models). During this phase atoms are no longer neither added nor removed. Initially, the optimization takes only into account the model structures, minimizing the difference between the projected positions of the new 3D model and those of the 2D models developed during the previous optimization phase. After this process has converged (see Figure~\ref{3D}a), the optimization is continued based on the error in simulated STEM-MAADF images as compared to the experimental ones. The convergence behaviour of this optimization phase is shown in Figure~\ref{3D}b. The model structures, simulated STEM-MAADF images and the error with respect to the experimental images are shown at three different stages of the process in Figure~\ref{3D}c. The yellow dashed line shows the area which is included in the model (edge atoms are excluded, but retained as background in the simulated image so that no discontinuity appears at the edge). Of course, the simulation of the first (untilted) model (View 1 in Figure~\ref{3D}c) fits perfectly to the corresponding experimental image (this was the starting configuration), but there is a high discrepancy between the experimental image and the simulation of the second (tilted) model (View 2). The situation improves quickly during the optimization process until at the end the difference between the experimental images and the simulated ones is dominated by noise.

In order to validate our approach and estimate its accuracy, we test it using a computationally obtained structure and simulated STEM images for realistic conditions including noise as expected for our experimental dose. For this purpose we use a defect configuration that was also observed experimentally, but with its 3D configuration obtained by energy minimization (see Methods). Since this defect contains two extra atoms compared to an ideal graphene lattice, it displays a significant out of plane distortion and hence is ideally suited for the validation. The reconstruction from the simulated data agrees with the original model with a mean out-of-plane deviation of 0.183~{\AA} and a mean in-plane deviation of 0.053~{\AA}. Details of this test are given in the supplementary information.  As expected, the out-of-plane error is larger than the in-plane error, since the effect of noise is amplified by the limited tilt angle between the two images. 

\begin{figure}
\center \includegraphics[width=1\textwidth]{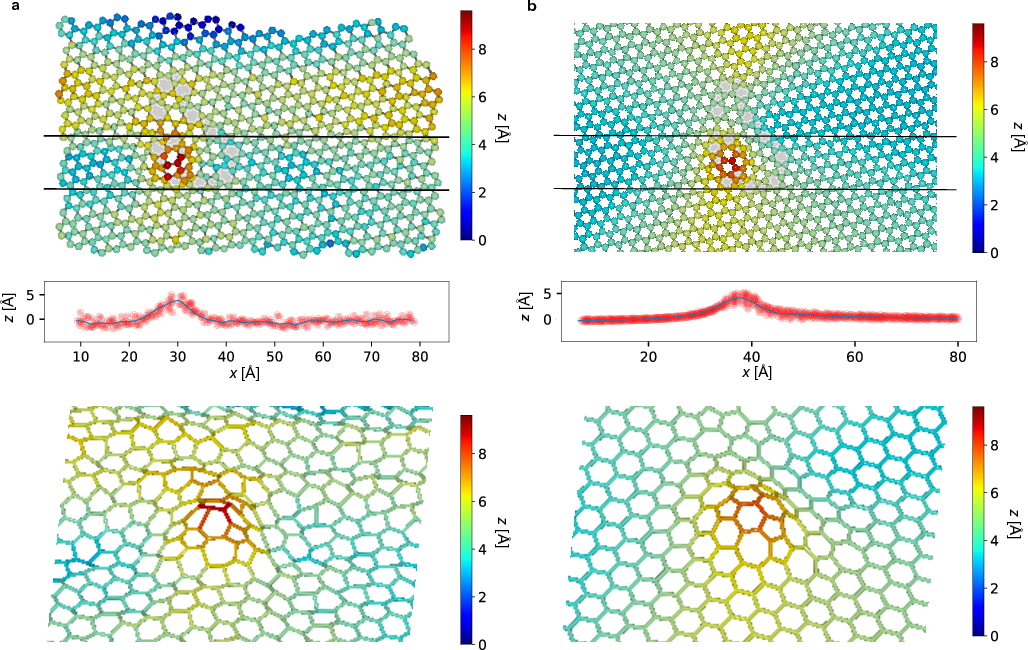}{}{}
    \caption{3D structure of an embedded defect with a rotated grain and an ad-dimer. (a) Structure obtained through experimental reconstruction from STEM-MAADF images. (b) Structure obtained computationally through energy minimization. The line profiles under the top views in both panels include all atoms between the two horizontal black lines. At the bottom, a perspective view of the structure is shown (only color coded bonds are shown).}
\label{mess} 
\end{figure}

Next, we use the two experimentally obtained STEM images of this structure. Figure \ref{mess} shows the experimental reconstruction (a) and for comparison the energy minimized structure (b). The top views, the lineprofiles and the side views show and excellent match between the two. The structure displays a particularly strong distortion around the two atoms that are furthest from the plane, and which can be identified as a carbon ad-dimer integrated into the defect.  Additionally, we tested our method with a small rotated grain (flower defect \cite{Park2010,Cockayne2011}), which is flat according to both, computational analysis (energy minimization) and 3D reconstruction from experimental data. For the latter structure, we could also calculate the out-of-plane standard deviation of the coordinates. In this case, the deviation is 0.31~{\AA}, which is slighly higher than the error obtained from the simulated data as discussed above. 

Since a total of four images could be obtained from one particular grain boundary (GB), we further test the method by performing multiple reconstructions from different pairs of images. All of the reconstructed GBs show very similar tendencies to an out-of-plane deformation at the defect locations (see supplementary information). The consistent results are another confirmation for the reliability of our method.

\begin{figure}
\center \includegraphics[height=0.6\textheight, keepaspectratio]{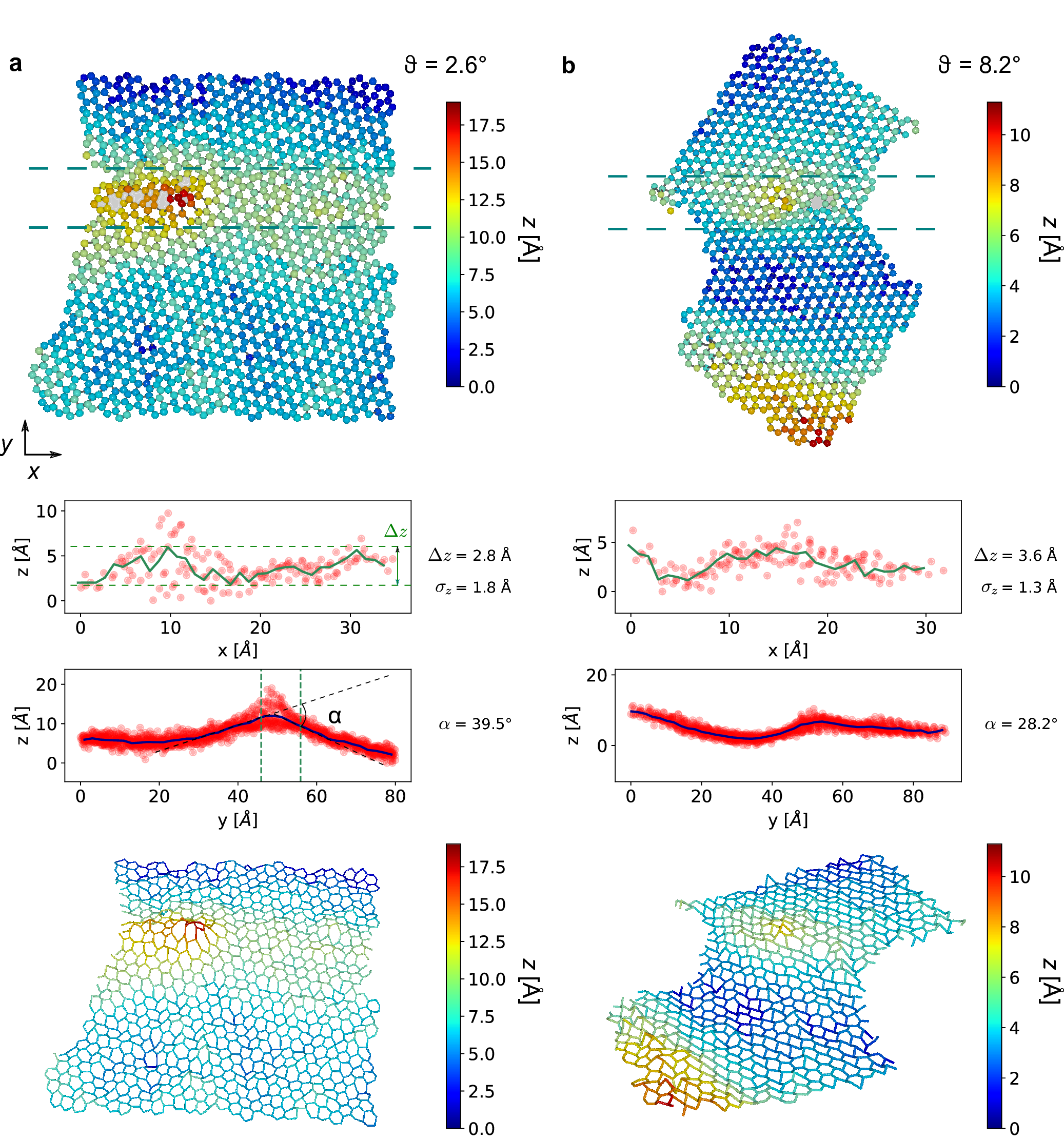}{}{}
    \caption{Atomic structures of two representative grain boundaries in graphene with small misorientation angles ($\theta$) of (a) ca. $2^\circ$ and (b) ca. $8^\circ$. From top to bottom, in each case: (1) Top view of the structure, with the z coordinate coded by color; (2) A side view of the atoms between the two green dashed lines, revealing height variations along the grain boundary (solid line shows an average); (3) A side view of the entire structure, viewed along the grain boundary in order to reveal the kink angle; (4) Perspective view of the structure (only bonds are shown, color coded for z position).  Also indicated in (a) are the definitions of the  maximum corrugation ($\Delta z$, peak to peak of the height variation along the grain boundary), the height variation ($\sigma_z$, standard deviation) and the kink angle ($\alpha$,  inclination between the two graphene sheets).}
\label{other_gbs} 
\end{figure}

\begin{figure}
\center \includegraphics[width=1\textwidth, height=1\textheight, keepaspectratio]{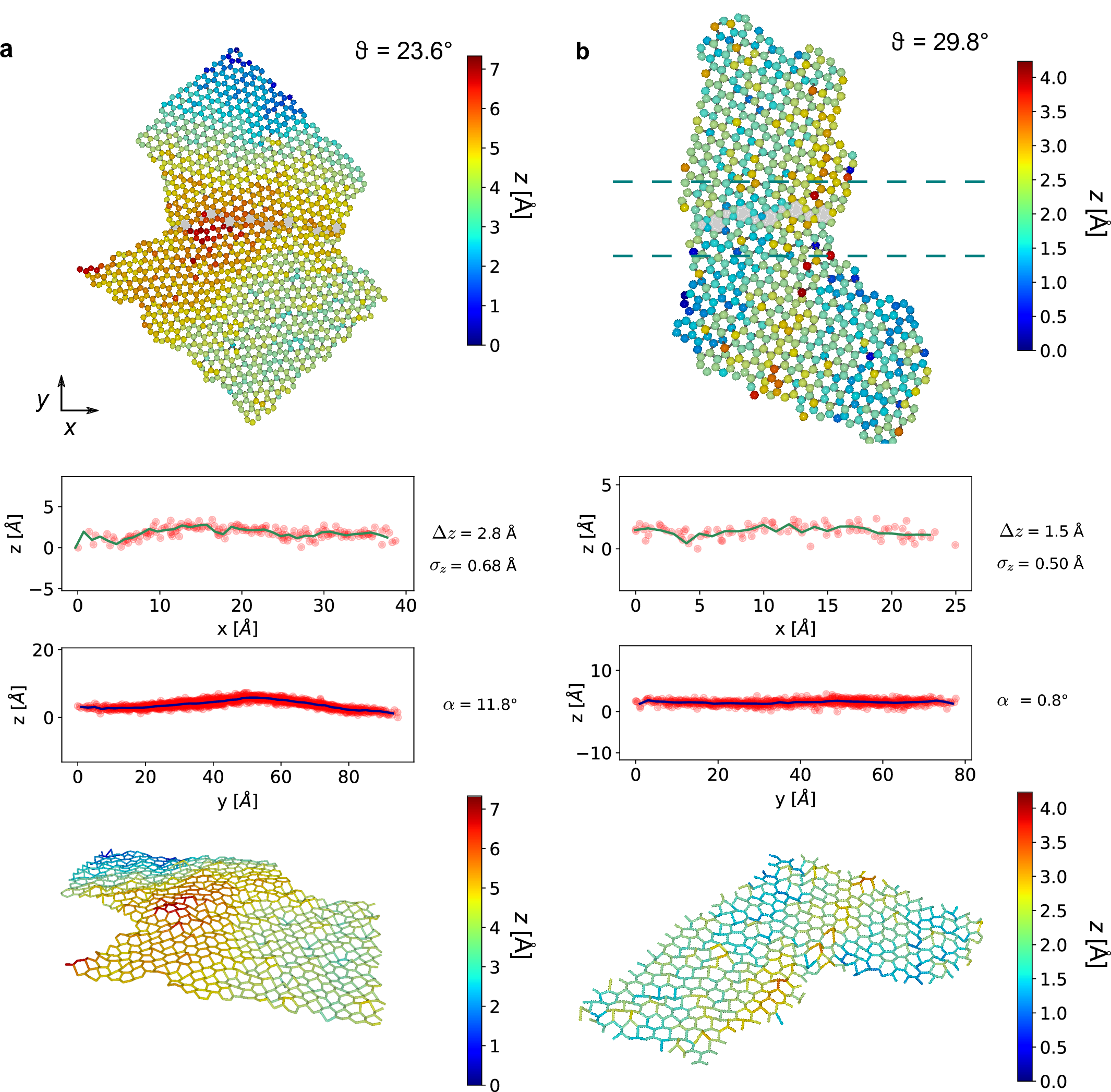}{}{}
    \caption{Atomic structures of two representative grain boundaries in graphene with large misorientation angles ($\theta$) of (a) ca. $23.6^\circ$ and (b) almost $30^\circ$. The data is displayed in the same way as in Figure \ref{other_gbs}a.}
\label{other_gbs2} 
\end{figure}


After establishing the reliability of the method for resolving the 3D structure, we move onto the analysis of graphene grain boundaries. GBs are a challenge for computational techniques, because they join together two crystalline grains with different orientations, hence they can neither trivially be incorporated into a periodic super cell required for most computational techniques nor can their effect on the surrounding graphene lattice be easily estimated with calculations using non-periodic structures since often only a short piece of a GB is visible in atomic resolution images. We present results for several GBs spanning misorientation angles $\theta \in [2^\circ,30^\circ]$. In Figure~\ref{other_gbs}, the 3D structure of two representative grain boundaries with small misorientation angles ($\theta= 2^\circ$ and $\theta= 8^\circ$, respectively) and in Figure~\ref{other_gbs2}, the 3D structure of two representative grain boundaries with large misorientation angles ($\theta= 24^\circ$ and $\theta= 30^\circ$, respectively) are shown. Three additional structures can be found in the supplementary information. For each structure, we show the top view colored based on the $z$-coordinate of each atom as well as two line profiles: one along the $y$-axis that contains all atoms in the structure and another along the $x$-axis that is limited to a narrow strip of atoms located within ca. 1~nm around the GB. From the line profiles, we also calculate the maximum corrugation ($\Delta z$), height variation ($\sigma_z$), defined as the standard deviation of the out-of-plane coordinate of the atoms in the structure from the mean value and the kink angle ($\alpha$) which is measured across the grain boundary. For an optical impression of the structure, we also show a perspective view of the bonds with the same color code as in the top view.

\begin{figure}
\center \includegraphics[width=\textwidth]{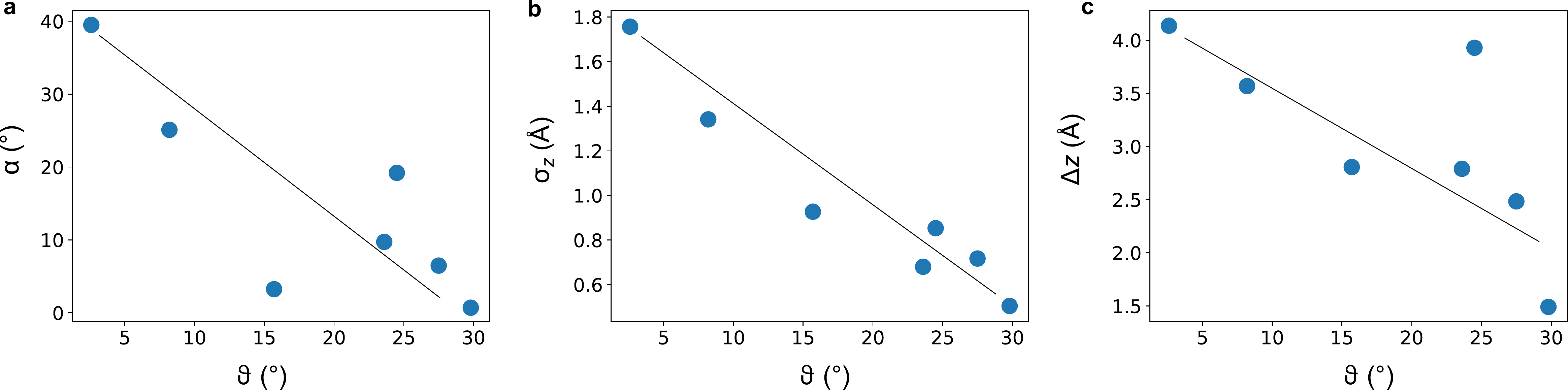}{}{}
\caption{Kink angle ($\alpha$), height variation ($\sigma_z$) and maximum corrugation ($\Delta z$) for different grain boundaries as functions of the misorientation angle ($\theta$). The lines are linear fits to the data, which serve as guides to the eye.}
\label{misor} 
\end{figure}

When each of the structural characteristics are plotted as a function of the misorientation angle ($\theta$) (Figure~\ref{misor}), it becomes clear that they all depend strongly on $\theta$. While the trend is most clear for the small- and high-angle grain boundaries, the intermediate data points display some scatter reflecting the large structural variability in these grain boundaries.  Remarkably, the lowest measured kink angle is only $\alpha \approx 0.7^\circ$ for a GB with $\theta \approx 30^\circ$, whereas the highest one is nearly $40^\circ$ for a GB with $\theta \approx 2.6^\circ$.  This variation is lower than what has been predicted based on density functional tight binding calculations~\cite{Malola2010}; the largest calculated angles were up to 85$^{\circ}$ with no clear correlation to the misorientation angle. This discrepancy is likely a consequence of the fact that the theoretical models were created by forcing two straight graphene edges to join, whereas during actual growth nothing prevents the carbon atoms from forming more meandering structures~\cite{Huang2011,Kim2011,Kurasch2012,Kotakoski2012} that help in reducing the stress at the GB. In another theoretical work~\cite{Yazyev2010}, it was 
predicted that small angle grain boundaries should show a pronounced tendency for buckling, whereas large angle grain boundaries tend to be flat. Although also this work was limited to straight GBs (and also formed of regular arrays of dislocations), this prediction is in good agreement with our experimental result. We indeed find that smaller $\theta$ predicts higher corrugation (up to $\Delta z \approx 4$~{\AA} and $\sigma_z \approx 1.8$~{\AA}), whereas the GBs with $\theta \sim 30^\circ$ tend to be significantly flatter (with  $\Delta z \approx 1.5$~{\AA} and $\sigma_z \approx 0.5$~{\AA}). These values reflect the fact that small-angle GBs contain isolated non-hexagonal rings as well as short segments where the hexagonal lattices of both grains are directly connected, leading to significant local strain that must be released through buckling~\cite{Yazyev2010,Liu2010}. This is particularly clear in Fig. \ref{other_gbs}a, where an essentially isolated dislocation core in a grain boundary with only $2.6^\circ$ misorientation angle is observed. 


In conclusion, we have demonstrated a new approach to determine the 3D structure of defective graphene at the atomic resolution from only two scanning transmission electron microscopy images taken at different sample tilts (with a respective difference of ca. $20^\circ$).  We first showed an embedded defect for which the results could be directly compared to the structure obtained through energy minimization. The comparison revealed excellent agreement, except for small local height variations due to noise in the experimentally obtained structure. We then applied the method to a set of grain boundary structures with misorientation angles nearly spanning the whole available range ($2.6-29.8^\circ$). The measured height variations at the boundaries reveal a strong correlation with the misorientation angle with lower angles resulting in stronger corrugation and larger kink angle (slope difference for the graphene grains on the different sides of the boundary). The largest measured kink angle was almost $40^\circ$ for a GB with $2.6^\circ$ misorientation. As far as we know, our results allow for the first time a direct comparison with theoretical predictions for the corrugation at grain boundaries. The measured kink angles are significantly smaller than the largest predicted ones~\cite{Malola2010}, probably due to artificial constraints in the theoretical models being different from the experimental reality. However, our results do qualitatively agree with the prediction that smaller misorientation leads to higher overall corrugation at the boundary~\cite{Yazyev2010}. Our results both open the way toward detailed study of the complete morphology of two-dimensional materials, including the often disregarded third dimension, and can already be used for tailoring graphene growth towards application utilizing the revealed differences in corrugations of polycrystalline samples with different misorientation angles between the graphene grains.

\section*{Methods}

{\bf Samples}---For our experiments, we studied chemical vapor deposition (CVD)
grown graphene. We used commercially available graphene on TEM
grids (Graphenea on Quantifoil R2/4), as well as self-grown CVD samples
transferred to Quantifoil 0.6/1 grids. In order to have a high defect density in our samples, we kept the growth temperature low ($T=960~^\circ C$), the flow rate high ($S_F=100~sccm$) and the annealing time as short as possible by starting the growth when the growth temperature is reached. As precursor, we used ethane which further increases the nucleation density.

{\bf Microscopy}---Scanning transmission electron
microscopy (STEM) experiments were conducted using a Nion UltraSTEM100, operated at 60 kV acceleration voltage. Typically, our atomic-resolution images were recorded with $512\times 512$~pixels for a field of view of 8~$nm$ and dwell time of 32~$\mu$s per pixel using the medium angle annular dark field (MAADF) detector.

{\bf Conjugent gradient energy minimization}---In order to study the strain adaptation in each of the defects, a supercell of pristine graphene with the size of 72$\times$62~nm consisisting of 173,000 atoms was created and the  defect structure is incorporated into this supercell.
LCBOP \cite{Los2003} was used as the long-range bond-order potential for carbon to describe the pair interactions. All calculations were performed with Large-scale Atomic/Molecular Massively Parallel Simulator code \cite{PLIMPTON19951,Plimpton2012}.
The total potential energy was minimized by relaxing atoms until the forces were below $10^{-3}$ eV/\AA\hspace{0.05cm} and the strain at the borders of the graphene flake was negligible (pressure below 1 atmosphere).

{\bf STEM image simulation}---Instead a multislice algorithm, which is typically used for quantitative STEM-simulations, we used a simplified method which works for single layer materials and is much faster. We approximate the potential of the 2D lattice by a zero-filled image with non-zero pixel values on the atomic positions. The simulation is obtained by convoluting this image with a (potentially aberrated) electron probe (see supplementary information for more details).

\section*{Supporting Information}
The Supporting Information contains details of the STEM simulation method, additional analysis of the 3D accuracy, additional grain boundary models and a complete set of raw and processed images. References \cite{Koch2002} and \cite{Kirkland2010a} are cited in the supplement.

\section*{Acknowledgments}

This work was supported by the European Research Council Starting Grant no. 336453-
PICOMAT. M.R.A.M, G.A. and J.K. acknowledge support from the Austrian Science Fund (FWF) through project I3181-N26.

\end{document}